\documentclass[aps,prc,groupedaddress,showpacs,nofootinbib,showkeys]{revtex4}

\usepackage{graphicx}
\usepackage{epsfig}
\usepackage{amsmath}
\newcommand{\az}{_{A,Z}}
\newcommand{\midtilde}{\raisebox{-0.25\baselineskip}{\textasciitilde}}
\usepackage{color}

\begin{document}

\title{Light clusters in nuclear matter: Excluded volume versus quantum many-body approaches}

\author{Matthias Hempel}
\email{matthias.hempel@unibas.ch}
\affiliation{Department of Physics, University of Basel, Klingelbergstrasse 82,
4056 Basel, Switzerland}

\author{J\"urgen Schaffner-Bielich}
\email{schaffner-bielich@thphys.uni-heidelberg.de}
\affiliation{Institut f\"ur Theoretische Physik, Ruprecht-Karls-Universit\"at,
Philosophenweg 16, 69120 Heidelberg, Germany}
 
\author{Stefan Typel}
\email{s.typel@gsi.de}
\affiliation{Excellence Cluster Universe, Technische Universit\"at M\"unchen,
Boltzmannstra{\ss}e 2, 85748 Garching, Germany \\ GSI Helmholtzzentrum f\"ur
Schwerionenforschung GmbH, Theorie, Planckstra{\ss}e 1, 64291 Darmstadt,
Germany}
 
\author{Gerd R\"opke}
\email{gerd.roepke@uni-rostock.de}
\affiliation{Institut f\"ur Physik, Universit\"at Rostock, Universit\"atsplatz
3, 18051 Rostock, Germany}


\begin{abstract}
The formation of clusters in nuclear matter is investigated, which occurs e.g.\
in low-energy heavy-ion collisions or core-collapse supernovae. In astrophysical applications, 
the excluded volume concept is commonly used for the description of light clusters.
Here we compare
a phenomenological excluded volume approach to two quantum many-body models, the
quantum statistical model and the generalized relativistic mean-field model. All
three models contain bound states of nuclei with mass number $A\leq4$. 
It is explored to which extent the complex medium effects can be mimicked by the
simpler excluded volume model, regarding the chemical composition and
thermodynamic variables. Furthermore, the role of heavy nuclei and excited
states is investigated by use of the excluded volume model. At temperatures of a
few MeV the excluded volume model gives a poor description of the medium effects
on the light clusters, but there the composition is actually dominated by heavy
nuclei. At larger temperatures there is a rather good agreement, whereas some
smaller differences and model dependencies remain.
\end{abstract}
\maketitle
\section{Introduction}
Clusters in nuclear matter play an important role in various physical systems.
Recently, it was shown that the behavior of the symmetry energy in low-energy
heavy-ion collisions can only be explained if the formation of clusters is
consistently taken into account \cite{natowitz10}. The appearance of nuclear
clusters in matter below saturation density leads to an increase of the binding
energy of symmetric nuclear matter, so that e.g.\ the symmetry energy gets a
finite value at zero density \cite{typel09}. Very recently there
was another attempt to connect predictions from different equations of state
(EOS) with experimental data from heavy-ion collisions \cite{qin2011}. 

Also in astrophysical environments the formation of clusters, respectively
nuclei, plays a crucial role. In the crust of cold neutron stars up to $8 \times
10^6$ g/cm$^3$ the ground state of matter is given by $^{56}$Fe
\cite{ruester06}, which is the nucleus with the lowest total energy (i.e.\
including rest masses) per nucleon. In general, in catalyzed cold neutron stars
only heavy nuclei appear and the fraction of light nuclei like deuterons or
alpha particles is vanishingly small. However, this is not the case for systems
at finite temperature. Larger temperatures in general favor the appearance of
lighter particles, due to the increased entropy. In core-collapse supernovae the
entropy per baryon typically ranges from 1 to 20 $k_B$ and the temperature ranges from
0 to 50 MeV\footnote{Throughout the article we use natural units, i.e.\
$\hbar=c=k_B=1$, where it is appropriate.}. In the shock heated matter there is
a significant contribution of light nuclei, whereas the fraction of heavy nuclei
is negligible. 

For astrophysical simulations of e.g.\ core-collapse supernovae or neutron
star mergers, equation of state tables are needed which cover the large domain
in baryon number density $10^{-10}$ fm$^{-3} < n_B < 1$ fm$^{-3}$, temperature
$0 < T < 100$ MeV, and proton fraction $0< Y_p< 0.6$ (see e.g.\ Ref.~\cite{fischer11}), with a typical resolution
summing up to more than one million grid points. Apart from the non-trivial
calculation of such tables, one has to encounter the conceptual challenge of a
consistent model for the EOS which is able to describe the formation of nuclear
clusters and heavy nuclei at low densities as well as the nuclear interactions
of uniform nuclear matter at densities above saturation density, with a smooth
transition between the different regimes. Due to these difficulties, at present
only a few EOS tables are available which are suitable for use in simulations of
core-collapse supernovae. In most simulations, the EOS of Lattimer and Swesty (LS) 
\cite{lattimer91} or of H.~Shen et al.\ (STOS) \cite{shen98, shen98_2} were used. 
In these two models, besides unbound nucleons and one representative heavy
nucleus, only alpha particles are considered. Furthermore, the dissolution of the
alpha particle at high densities is only modeled by an excluded volume
mechanism. Recently, two new EOS tables were published by G.~Shen et al.\
\cite{shen2011a,shen2011b}, in which the interaction of alpha particles with the
nucleons is described more realistically by using the virial expansion
\cite{horowitz06b} at low densities. In Ref.~\cite{hempel10}, Hempel and
Schaffner-Bielich (HS) developed a new EOS model and later calculated new EOS
tables which have been used in simulations of core-collapse supernovae in
Ref.~\cite{hempel11}. The EOS of HS is based on a nuclear statistical
equilibrium model with interactions of the unbound nucleons and excluded volume
effects for the nuclei. Important for our purposes here, it also contains all
possible light nuclei. 

In the simulations of Ref.~\cite{hempel11} it was found, in comparison with the
LS and STOS EOS, that the alpha particle only poorly describes the entire
distribution of all possible light clusters. Especially deuterons and tritons
appear with large abundances in the shock heated matter behind the standing
bounce shock and can even surmount the alpha-particle fraction. Similar results
were found in the study of Sumiyoshi and R\"opke \cite{sumiyoshi08}  before,
where light nuclei were included in a core-collapse supernova simulation based
on the quantum statistical model of R\"opke \cite{roepke09}, but only in
post-processing of the simulation data. These two studies, and many other
theoretical investigations \cite{oconnor07, arcones08, typel09, heckel09}, show
that additional light nuclei appear with large abundances in supernova matter.
There is also experimental evidence for this statement: in the previously
mentioned heavy-ion experiment similar conditions as in a core-collapse
supernova are obtained, and the experimental results cannot be explained without
taking light clusters into consideration.

However, there are no simulations of core-collapse supernovae, which
consistently take into account all light clusters in the EOS and their
interaction with leptons in weak reactions. There exist some interesting
exploratory studies. In Ref.~\cite{arcones08} the influence of light nuclei on
neutrino-driven supernova outflows was investigated and a significant change in
the energy of the emitted antineutrinos was found. As another example, in
Ref.~\cite{oconnor07} it was shown, that mass-three nuclei contribute
significantly to the neutrino energy loss for $T\geq4$ MeV. In the simulations
with the EOS of HS \cite{hempel11}, weak reactions on the additional light
nuclei were described in a simplified way by treating them as alpha particles.
The authors of this study conclude that the presence of additional light nuclei
might contribute to the neutrino heating in the early post-bounce phase, whereas
in the later evolution the effect on cooling might be more important. Still, the
question remains to be answered how important the effect of the light cluster
distribution will be on the supernova dynamics.

In the present article we compare the statistical excluded volume model (ExV) of
HS \cite{hempel10} with two different quantum many-body models, which also
include the bound states of light nuclei, the generalized relativistic
mean-field (gRMF) model of Typel et al.\ \cite{typel09}, and the quantum
statistical (QS) model of R\"opke \cite{roepke09,roepke11a}. The two quantum
many-body models give a detailed description of the medium effects on light
nuclei. On the other hand, it is much more demanding to include heavy nuclei in
the quantum many-body models than in the excluded volume
description. This is necessary if one wants to cover the
aforementioned parameter range relevant for core-collapse supernova simulations.
So far this has been done only for selected thermodynamic conditions by use of
the single nucleus approximation \cite{typel09}. In this approach the full distribution of
heavy nuclei is replaced by a single representative nucleus. For the ExV
model, EOS tables are already available online, which include the full
distribution of light and heavy nuclei, and cover a broad range in density,
temperature, and
asymmetry.\footnote{\texttt{See http://phys-merger.physik.unibas.ch/\midtilde
hempel/eos.html}\label{eospage}.} Here, we want to judge the reliability of the
excluded volume approach which is also used in many other EOS models and
astrophysical applications and want to see to what extent it can mimic the
medium effects of a more microscopic quantum statistical description. 
We will discuss results for symmetric nuclear matter at a given temperature $T$.
Thus the state of the system is defined by $(T,n_B,Y_p)$, with the baryon number
density being $n_B$ and the total proton fraction $Y_p=0.5$. In the comparison of the
different models we also want to identify the parameter range where heavy nuclei
give the dominant contribution to the composition. Furthermore, we will also
address the role of excited states of hot nuclei. The article is structured as
follows. In Sec.~\ref{sec_nse} we briefly present the excluded volume model, and
in Sec.~\ref{sec_qmb} we present the quantum many-body models. In Sec.~\ref{sec_results} we
will compare the results of the three different approaches, before we conclude
in Sec.~\ref{sec_conc}.

\section{Excluded Volume Model}
\label{sec_nse}
The ExV model consists of an ensemble of nucleons and nuclei in nuclear
statistical equilibrium, where interactions between the nucleons and excluded
volume corrections for the nuclei are implemented. Here we will only give a
brief summary of this model; all details can be found in Ref.~\cite{hempel10}. 
For the unbound interacting nucleons (neutrons and protons) a density-dependent
relativistic mean-field (RMF) model \cite{typel05} is applied. Its Lagrangian is
based on the exchange of the isoscalar-scalar $\sigma$, the isoscalar-vector
$\omega$ and the isovector-vector $\rho$ mesons between nucleons. The coupling
strengths are assumed to be density dependent. The free parameters of the
Lagrangian, the masses of the mesons and the parameters of the couplings, have
to be determined by fits to experimental data. Here we apply the parameter set
DD2 \cite{typel09} in order to use the same approach for the nucleons as the QS
and gRMF model in the comparison of the results. The parametrization is based on
the same nuclear input as the parameter set DD \cite{typel05}, but experimental
nucleon masses are used. Previous tabulations of the ExV model were based on
different RMF parameter sets with nonlinear self-interactions of the mesons.

In the ExV model nuclei are treated as non-relativistic classical particles with
Maxwell-Boltzmann statistics. We preferably use experimentally measured masses
for the description of nuclei. For that we take the nuclear data from the atomic
mass table 2003 from Audi et al.\ \cite{AudiWapstra} whenever
possible. For nuclei with an experimentally unknown mass we use the results of the
finite-range droplet model (FRDM) in the form of a nuclear mass table
\cite{Moller97}. Nuclei behind the neutron drip line are not considered. Below
we will show results for two different cases: first, if only the most important
light clusters with mass number $A\leq 4$ are considered, and then, if all
nuclei with available binding energies are taken into account.

For the screening of the Coulomb field of the nuclei by the background of
electrons we use the most basic approximation: each nucleus with mass number $A$
and charge number $Z$ is placed in the center of a spherical Wigner-Seitz (WS)
cell with uniform electron density. This leads to the Coulomb correction of the
energy of the nucleus: 
\begin{equation}
 E^{\textrm {Coul}}\az=- \frac35 \frac{Z^2 \alpha}{R_A}\left(\frac32 x- \frac12
x^3\right), \; x=\left(\frac{n_B}{2n_B^0}\frac{A}{Z}\right)^{1/3} 
\label{nse_eq_ecoul}
\end{equation}
where $R_A$ is the radius of the nucleus with mass number $A$. Here it is
assumed that nuclei are uniformly charged spheres at saturation density
$n_B^0=0.149$~fm$^{-3}$, the value of the DD2 parametrization \cite{typel09},
so that $R_A=(3A/4 \pi n_B^0)^{1/3}$. More elaborated approaches for the Coulomb
energy of a multi-component plasma at finite temperature can, e.g., be found in
Refs.\ \cite{nadyozhin05,potekhin09,potekhin10}. However, the Coulomb energy
becomes only important for heavy nuclei at low temperatures, which is not the
main subject of the present investigation. We will even neglect the Coulomb
energy for the cases where we only consider the light nuclei with $A\leq 4$.

To take into account excited states of the nuclei, we use a temperature-dependent degeneracy function, i.e.~an internal partition function, $g_{AZ}(T)$.
It represents the sum over all excited states of a hot nucleus. We choose the
simple semi-empirical expression from Ref.~\cite{fai82}:
\begin{equation}
g_{AZ}(T)=g_{AZ}^0 +\frac{0.2}{A^{5/3}\textrm{MeV}}\int_0^{E_{\rm max}} dE^*
e^{-E^*/T}\exp\left(\sqrt{2 a(A) E^*}\right), \;  a(A)=\frac A 8 (1-0.8A^{-1/3})
\rm {~MeV} ^{-1} \label{eq_gaz}
\end{equation}
with $g_{AZ}^0$ denoting the spin degeneracy of the ground state. For the
maximal excitation energy $E_{\rm max}$ we choose the binding energy of the
nucleus (see Ref.~\cite{roepke84}). This means we only consider excited states
which are still bound. Later we will show results for two cases: Either we
consider only the ground-state spin degeneracy or we apply the temperature-dependent internal partition function. We did not include any experimentally
known states, because the coverage is rather incomplete and we preferred to have
a uniform description for all nuclei. We checked that the results did not change
significantly if only the known excited states were used for the light nuclei.

In our description we distinguish between nuclei and the surrounding interacting
nucleons, and we still have to specify how the system is composed of the
different particles. Let us denote the number density of neutrons and
protons by $n_n$ and $n_p$, respectively, and of the nucleus $(A,Z)$ by $n_{A,Z}$.
Thus here and in the following we require $A\geq 2$. The total baryon number
density and the total proton number density are then given by:
\begin{eqnarray}
n_B&=& n_n+n_p+\sum \az A~n\az \; ,
\end{eqnarray}
\begin{eqnarray}
n_BY_p&=& n_p+\sum \az Z n\az \; .
\end{eqnarray}
For nuclei we will apply the following phenomenological description: All baryons
(nucleons in nuclei or unbound nucleons) are treated as hard spheres with the
volume $1/n_B^0$ so that the nuclei are uniform hard spheres at saturation
density of volume $V_{A}=A/n_B^0$. Next, a nucleus must not overlap with any
other baryon (nuclei or unbound nucleons). Thus the volume in which the nuclei
can move is not the total volume of the system, but only the volume which is not
filled by baryons. With these assumptions the free volume fraction is given by:
\begin{eqnarray}
\kappa&=&1-n_B/n_B^0 \label{nse_eq_kappa} \; .
\end{eqnarray}
For the unbound nucleons we use a different description, because the
interactions among them are already included in the RMF model. For unbound
nucleons we only assume that they are not allowed to be situated inside of
nuclei. Therefore we introduce the volume fraction which is not filled by
nuclei, i.e.\ the filling factor of the nucleons:
\begin{equation}
\xi=1-\sum \az A~n\az/n_B^0 \; .
\end{equation}
Next we introduce the number densities of nucleons outside of nuclei, $n_n'$ and $n_p'$, i.e.~the number of neutrons and protons, respectively, per
volume which is not filled by nuclei. They are related to the densities $n_n$
and $n_p$ which are given by the number of neutrons and protons, respectively, per
total volume by:
\begin{eqnarray}
n_n'&=&n_n/\xi \nonumber \\
n_p'&=&n_p/\xi \; .
\end{eqnarray}

With these assumptions one derives the following free-energy density:
\begin{eqnarray}
f&=&\sum \az f\az^0(T,n\az)+\sum \az n\az E^{\rm Coul}_{A,Z} +\xi f_{\rm
nuc}^0(T,n'_n,n'_p)-T\sum \az n\az \mathrm{ln}(\kappa)\; . \label{eq_f}
\end{eqnarray}
The local number densities $n_n'$ and $n_p'$ set the RMF contribution of the
nucleons $f^0_{\rm nuc}$, which is weighted with their filling factor $\xi$.
This weighting can be expected, as the free energy is an extensive quantity. The
first term in Eq.~(\ref{eq_f}) 
\begin{eqnarray}
f\az^0&=& n\az\left\{M\az-T-T\mathrm{ln}\left[\frac{g\az(T)}{n
\az}\left(\frac{M\az T}{2\pi}\right)^{3/2}\right]\right\}
\end{eqnarray} 
is the ideal gas expression for the free-energy density of the nuclei. The
Coulomb free energy of the nuclei appears in addition. The last term in
Eq.~(\ref{eq_f}) is the direct contribution from the excluded volume. Because of
this term, as long as nuclei are present, the free-energy density goes to
infinity when approaching saturation density ($\kappa \rightarrow 0$). Thus
nuclei will always disappear before saturation density is reached. If no nuclei
are present, we get the unmodified RMF description, as it should be. The
excluded volume correction of the nuclei represents a hard-core repulsion of the
nuclei at large densities close to saturation density. Similarly, the
modification of the free energy of the unbound nucleons is purely geometric and
just describes that the nucleons fill only a fraction of the total volume. 

Chemical equilibrium between nuclei and nucleons leads to
\begin{eqnarray}
&&n \az=\kappa~g\az(T)\left(\frac{M\az
T}{2\pi}\right)^{3/2}\exp\left(\frac{(A-Z)\mu_{n}^0+Z\mu_{p}^0-M\az-E^{\rm
Coul}\az-p^0_{\rm nuc}V_{A}}T\right) \; , \label{nse_eq_naz}
\end{eqnarray}
where $V_{A}=A/n_B^0$ is the volume of the nucleus. $\mu_n^0$ and $\mu_p^0$, are
the chemical potentials of the neutrons and protons, respectively, and the
symbol $p^0_{\rm nuc}$ denotes their summed pressure. These quantities are
calculated with the relativistic mean-field model for $n_n'$ and $n_p'$.
Explicit expressions are given e.g.\ in Ref.~\cite{typel09}. All thermodynamic
quantities can be derived consistently in the standard manner from the free
energy; the results are given in Ref.~\cite{hempel10}. We note that the
presented approach for the excluded volume corrections is thermodynamically
fully consistent. It is also in agreement with the work of Yudin \cite{yudin11}, who studied excluded volume schemes in a rather general form. 

$\mu_n^0$, $\mu_p^0$, and $p^0_{\rm nuc}$ contain the RMF interactions of the
nucleons. As they appear in Eq.~(\ref{nse_eq_naz}) the interactions of the free
nucleons are coupled to the contribution of the nuclei. Nuclei are influenced by
the mean field of the nucleons. However, in contrast to the generalized RMF
model of Typel et al.~\cite{typel09}, which we will present in the next section,
the bound nucleons do not contribute to the source term of the meson fields, so
nuclei do not give a contribution to the mean field. Furthermore, the
mutual counteracting in-medium self-energy and the Pauli blocking shifts of the
light clusters do not appear. In the ExV model the Mott effect and the
dissolution of clusters at large densities is mimicked only by the excluded
volume corrections.

One can use also a different value than saturation density for
the nucleon density inside of nuclei. It can be seen as a free parameter of the
model which determines the strength of the excluded volume effects. This
parameter sets the threshold density above which nuclei cannot exist any more.
However, we found that other values do not give an overall better agreement with
the quantum models. Thus, here we only consider the most intuitive value
$n_B^0$, i.e.\ the saturation density of the DD2 parametrization.

\section{Quantum Many-Body Models}
\label{sec_qmb}
The generalized relativistic mean-field (gRMF) model has been introduced in
Ref.~\cite{typel09}. The same parametrization DD2 is applied as for the ExV
model. In the gRMF model, in addition to the nucleons, the light clusters are
included as quasiparticles which contribute as sources for the meson fields.
Like the nucleons, also the light clusters get a mean-field self-energy leading
to a reduced effective mass and medium shifts of the chemical potentials.
However, the light clusters are composite particles of nucleons. Thus, at large
densities the light clusters do not behave as free quasiparticles, but are influenced by the
filled Fermi sea of nucleons. This effect is called Pauli blocking and leads to
a shift in the binding energies which cannot be described by the gRMF model
itself. It is included as a density-dependent part of the nuclear masses, which
is taken from the quantum statistical (QS) model in parameterized form. 

The QS model is described in detail in Refs.~\cite{roepke09,roepke11a}. It is
based on the thermodynamic Green's-function method and uses an effective
nucleon-nucleon interaction. Effects of the correlated medium such as Pauli
blocking, Bose enhancement, and self-energy are taken into account, leading,
e.g., to the merging of bound states with the continuum of scattering states with
increasing density (the Mott effect). The nucleon self-energies in the QS model are
evaluated with the RMF model, neglecting the effect of cluster formation on the
mean fields. Then the medium modifications can be determined, such as the mass
shift and the Mott densities, where the clusters get dissolved.

As an example, the alpha particle is considered. From a Green's-function
approach, we obtain the in-medium four-particle wave equation
\begin{eqnarray}
&& \left[E^{\rm qu}(1)+E^{\rm qu}(2)+E^{\rm qu}(3)+E^{\rm qu}(4)-E^{\rm
qu}_\alpha(P)\right] \psi_{\alpha,P}(1,2,3,4) \nonumber \\ &&+ \sum_{p_1'p_2'}
[1-\tilde{f}(1)-\tilde{f}(2)] V(1,2;1',2')  \psi_{\alpha,P}(1',2',3,4)
 +... ({\rm permutations})= 0,
\end{eqnarray}
where five additional interaction terms obtained from different pairings are not
given explicitly. Similar equations hold for $A=2,3$. The single-particle
energies $E^{\rm qu}(1)$ contain the self-energy shifts, given, e.g., by the RMF
approach. The Pauli blocking contains the phase-space occupation $\tilde{f}(1)$.
If we neglect correlations in the surroundings, the actual phase-space
occupation (momentum distribution) is replaced by a Fermi function, with a
chemical potential that fits the nucleon densities.

The binding-energy shift of the alpha particle
\begin{equation}
\Delta E^{\rm qu}_\alpha(T, n_B, Y_p, P)= E^{\rm
  qu}_\alpha(P)-E^{(0)}_\alpha 
= \Delta E^{\rm SE}_\alpha(P)+ \Delta E^{\rm Coulomb}_\alpha(P)+\Delta E^{\rm
Pauli}_\alpha(P)
\end{equation}
with respect to the energy in the vacuum $E_{\alpha}^{(0)}$, contains the
self-energy shift due to the single-nucleon quasiparticle energies $\Delta
E^{\rm SE}_\alpha$, the Coulomb shift $\Delta E^{\rm Coulomb}_\alpha$, and the
Pauli blocking shift $\Delta E^{\rm Pauli}_\alpha$. The Pauli blocking shift is
relevant for the decrease of the binding energy of nuclei with increasing
density. In general, the total binding-energy shift depends on the center of mass
momentum $P$ of the nucleus with respect to the medium.

A parametrization of the shifts of the in-medium binding energies is given in
Ref.~\cite{typel09}. We use here the more recent parametrization
\cite{roepke11a} for the QS calculations. The results are nearly identical with
the results of the QS calculation shown in Ref.~\cite{typel09}. To evaluate the
composition, in the QS calculation also the continuum contributions have been
included. In this way the virial EOS is reproduced in the low-density limit. For
the single-nucleon quasiparticle shift, the DD2 parametrization was applied to
get the same behavior near saturation density as in the ExV and gRMF
calculations.

\section{Comparison of results}
\label{sec_results}
In the following we will compare the results for the three models ExV, gRMF, and
QS, for symmetric nuclear matter at three different temperatures. For this
comparison we will first only consider nucleons and light clusters with $A\le
4$, which are also used in Ref.~\cite{typel09}, i.e.\ neutrons, protons,
deuterons $^2$H, tritons $^3$H, helions $^3$He, and alpha particles $^4$He. To
investigate the role of heavier clusters, we will then include all available
nuclei in the ExV model. Finally we will study the role of excited states, by
also considering the internal partition function in ExV as presented in
Sec.~\ref{sec_nse}.

\begin{figure*}
\begin{center}
\includegraphics[width=0.75\columnwidth]{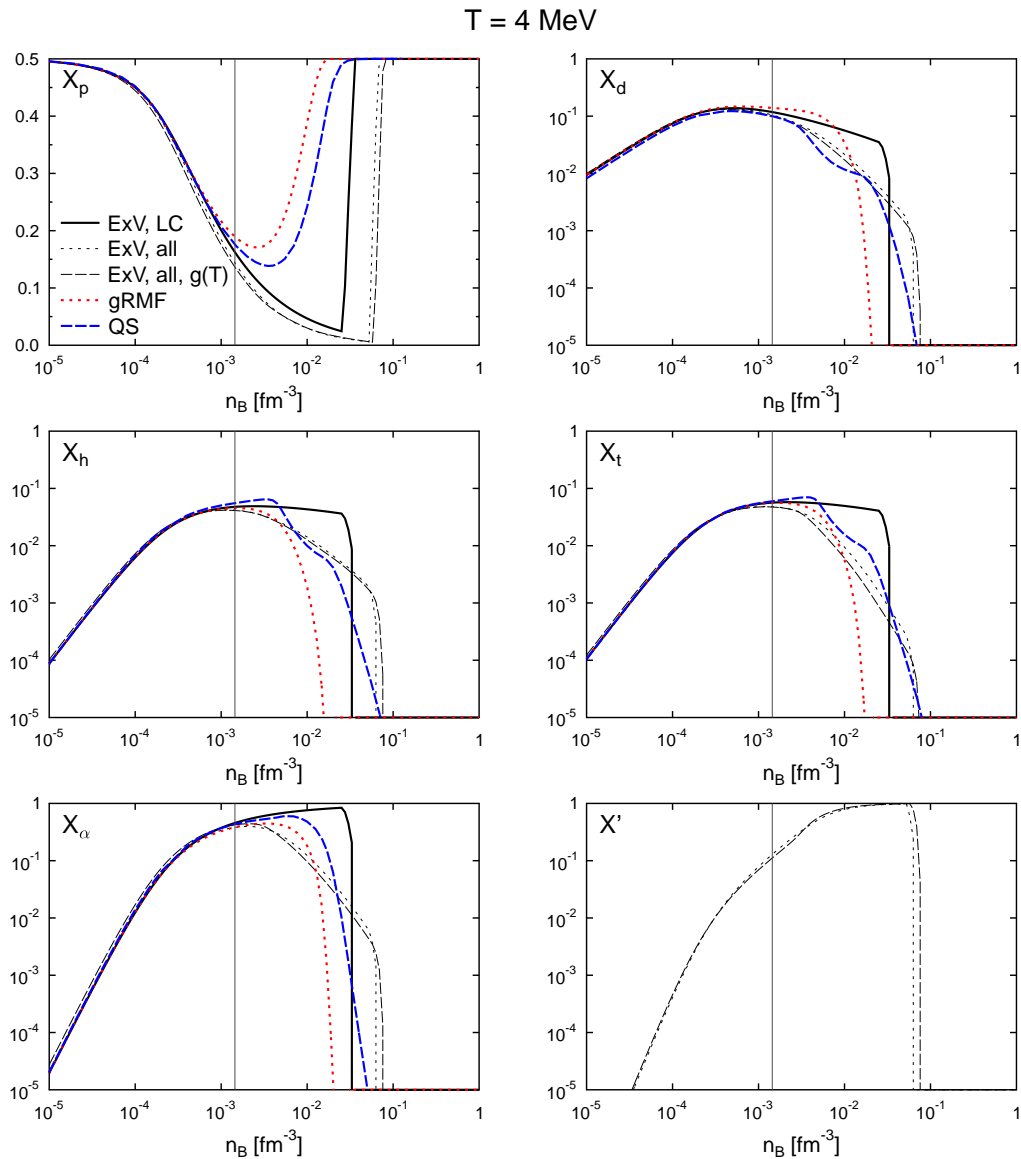}
\caption{\label{fig_comp1}The mass fractions of protons, deuterons, helions,
tritons, and alphas, and the mass fraction $X'$, defined by Eq.\ (\ref{eq_xs}),
for symmetric nuclear matter at $T=4$ MeV. The results of the generalized
relativistic mean-field model gRMF and of the quantum statistical model QS from
Ref.~\cite{typel09} are compared to the excluded volume model ExV
\cite{hempel10}. ``ExV, LC'' shows the results if only the same light clusters
with $A\leq 4$ as in Ref.~\cite{typel09} are considered and no excited states
are taken into account. For ``ExV, all'' all available nuclei are used. ``ExV,
all, g(T)'' also takes all available nuclei into account, but this time with the
internal partition function of excited states. The vertical gray lines show the
density where the fraction $X'> 0.1$ for ``ExV, all, g(T).''}
\end{center}
\end{figure*}

Figure \ref{fig_comp1} shows the comparison of the composition for $T=4$~MeV.
Let us first focus on the calculations without heavy nuclei, i.e.\ the thick
lines. The results with heavy nuclei (thin dashed and dotted lines) will be
discussed later. We depict the mass fractions $X_{i}=A_i n_{i}/n_B$, $i \in
\{p,d,h,t,\alpha\}$, of protons, deuterons, helions, tritons, and alphas. We
note that the fraction of tritons $X_t$ is almost equal to the helion fraction
$X_h$, because they are isospin partners, and we are investigating symmetric
nuclear matter. The only differences arise due to the electron screening of the
Coulomb interaction and the mass difference. Similarly, the mass fraction of
unbound neutrons (which is not shown) is almost equal to the unbound proton
fraction $X_p$. Up to $n_B\sim 10^{-3}$~fm$^{-3}$ the predictions of the
different models for the composition agree with each other. Pronounced
differences occur at larger densities. In the QS and gRMF model above the Mott
densities, the light clusters start to dissolve due to the Pauli blocking. The
binding energies of the light clusters are reduced gradually with density, which
leads to an increasing fraction of unbound protons above $n_B\sim 5 \times
10^{-3}$~fm$^{-3}$. Conversely, in the ExV model, the proton fraction decreases
and the total fraction of light clusters increases until $n_B \sim
0.036$~fm$^{-3}$, where a sudden turnover in the composition appears. The two
quantum many-body models do not show this behavior. They exhibit a more
continuous change of the particle fractions and are rather similar. Still they
have different features in detail, like e.g.\ in the QS model the sinous
behavior of some of the mass fractions and larger densities at which the
clusters disappear.

\begin{figure*}
\begin{center}
\includegraphics[width=0.75\columnwidth]{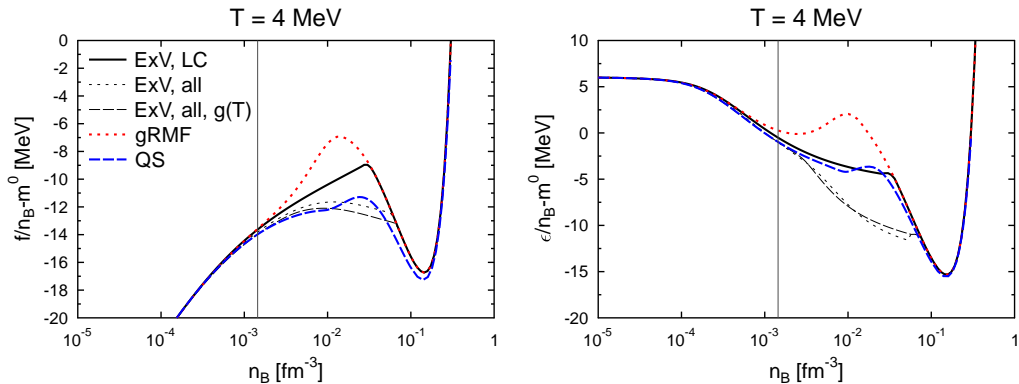}
\caption{\label{fig_thermo1}The free energy per baryon and internal energy per
baryon with respect to the averaged rest mass of neutrons and protons at $T=4$
MeV. Otherwise, notation is as in Fig.~\ref{fig_comp1}.}
\end{center}
\end{figure*}

The excluded volume approach gives a crude representation of the  cluster
dissolution at this moderate temperature. However, it is enlightening to study
the results of the ExV model if heavy nuclei are taken into account,
depicted by the thin black lines, where the dashed lines include excited states
and the dotted do not. At this moderate temperature excited states have only a
small impact on the composition. In the lower right panel of
Fig.~\ref{fig_comp1} we show
the sum of the mass fraction
\begin{equation}
X'=1-X_n-X_p-X_d-X_h-X_t-X_\alpha  \label{eq_xs}
\end{equation}
of all nuclei which are not included in the quantum many-body models (i.e.\
heavy nuclei with $A > 4$), but in the ExV model with binding energies using the
data base of Audi et al.\ \cite{AudiWapstra} and the FRDM mass
table \cite{Moller97}. One sees that light clusters are actually replaced by
heavy nuclei already at intermediate densities. $X'$ exceeds 0.1 for densities
larger than $1.4 \times 10^{-3}$ fm$^{-3}$, shown by the vertical lines in
Figs.~\ref{fig_comp1} and \ref{fig_thermo1}. $X'$ gets even larger than 0.9 for
densities above $0.016$ fm$^{-3}$. Thus the differences which are seen at such
large densities in the comparison for $A\leq4$ are not very significant, because
the composition is dominated by heavy nuclei there. The fraction of light
clusters is reduced considerably by the appearance of heavy nuclei, before the
Mott densities are reached. 

For helions and tritons the heavy nuclei lead to an interesting effect. At low
densities, where no heavy nuclei exist, the triton fraction is larger than the
the helion fraction because of the larger binding energy. However, at densities
larger than $ \sim 10^{-3}$ fm$^{-3}$, when the heavy nuclei appear, the triton
fraction decreases faster than the helion fraction. Heavy nuclei tend to be
asymmetric because of the Coulomb energy. When they give the main contribution
to the composition, it is favorable to absorb neutrons from light clusters. We
note that the same effect occurs for the unbound neutrons, which is reduced
compared to the unbound proton fraction. 

In Fig.~\ref{fig_thermo1} we show the free energy per baryon $f/n_B$ and the
internal energy per baryon $\epsilon/n_B$ for the different calculations. We
subtract the rest mass $m^0=1/2(m_n+m_p)$, corresponding to a proton fraction of
0.5. The free-energy density $f$ of the ExV model is specified in
Eq.~(\ref{eq_f}), and the internal energy density $\epsilon$ is given in
Ref.~\cite{hempel10}. The corresponding detailed expressions for the quantities
of the quantum many-body models can be found in Ref.~\cite{typel09}. First we
discuss the results of Fig.~\ref{fig_thermo1}, for which only light nuclei are
taken into account. Up to the density of $n_B\sim 10^{-3}$ fm$^{-3}$, where the
composition of the three different models still agrees, also the displayed
thermodynamic variables behave still similarly. At larger densities, even though
there is a better agreement for the composition between the QS and the gRMF
model, the internal energy of QS is more similar to the ExV model. One reason
for the difference between the QS and gRMF models is that the back reaction of
the change in the binding energies on the mean fields, mediated by additional
rearrangement contributions, is considered in the latter approach. In the QS
calculation, the self-energy shifts are taken from the RMF calculation in a
parametrized form for pure neutron-proton matter. The free energy of the ExV
model without heavy nuclei remains between the two quantum many-body models. One
can conclude, that a similar behavior of the composition does not imply in
general that other thermodynamic quantities also behave similarly.

The internal energy and free energy of the ExV model change significantly, if
heavy nuclei are included, whereas excited states play only a minor role.
Deviations due to heavy nuclei appear above $n_B \sim 10^{-3}$ fm$^{-3}$ and
become as large as the differences between the QS and gRMF models. It is
expected, that also the predictions of the QS and gRMF models for the internal
and free energies will be modified considerably if heavy nuclei are taken into
consideration in these approaches. However, if we restrict the comparison of the
three models to densities below the value shown by the vertical lines, there is
an excellent agreement of all three models. When uniform nuclear matter is
reached, the three EOSs give identical results, as they are built from the same
RMF parametrization DD2. 

\begin{figure*}
\begin{center}
\includegraphics[width=0.75\columnwidth]{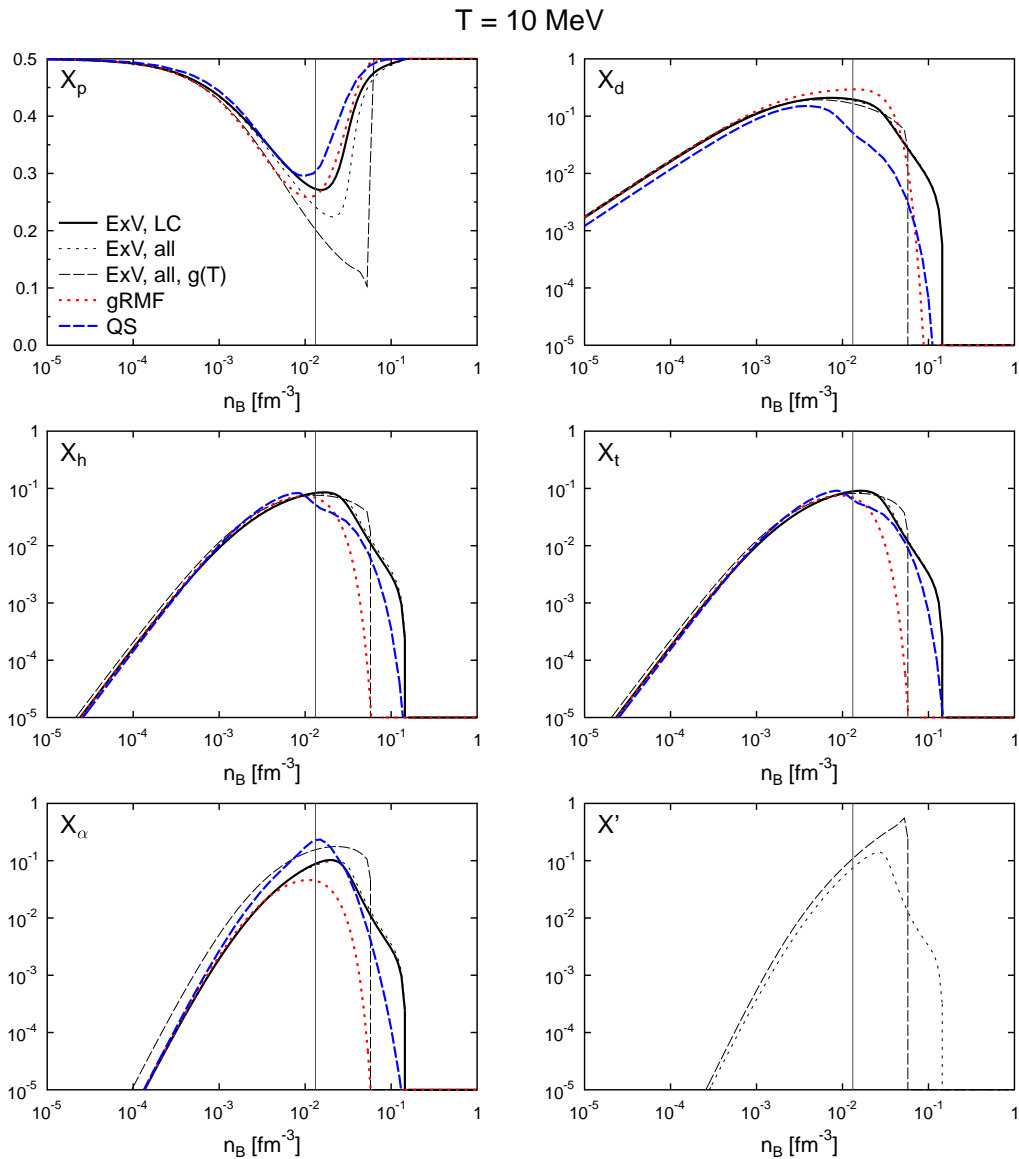}
\caption{\label{fig_comp2}As in Fig.~\ref{fig_comp1}, but now for $T=10$ MeV.}
\end{center}
\end{figure*}
 
Figure \ref{fig_comp2} shows the composition of matter for the three models at
$T=10$ MeV. Again, we first discuss the results where only light nuclei with
$A\leq4$ are considered. There is no sudden turnover of the composition in the
ExV model any more, but the clusters dissolve in a continuous way. At this
temperature the composition of the ExV model agrees much better with the two
quantum models. The maximum deuteron and alpha-particle fractions lie between
the results of gRMF and QS, and the maximum helion and triton fractions are almost
the same as in the QS model. The total cluster fraction of ExV is always close
to the two other models, as can be seen from the proton fraction. The densities
at which the light clusters disappear completely are slightly larger in the ExV
model. In Fig.~\ref{fig_comp2} one sees clearly that the deuteron fraction of QS
is slightly reduced, even at very low densities. This is due to the more
elaborated treatment of the deuterons in the QS model, in which its continuum
contributions are correctly subtracted. Overall the differences of the two
quantum models are of similar sizez as the differences to the ExV model. Thus we
can conclude that the ExV model mimics the quantum medium effects reasonably
well at $T=10$ MeV. We explain the better agreement at large temperatures by the
following aspects. First, the unbound nucleon density is in general larger at
larger temperatures, and clusters appear with lower fractions. This is a trivial
reason for the closer similarity. Second, the excluded volume corrections give a
contribution to the free-energy density proportional to $T \ln(\kappa)$ [see
Eq.~(\ref{eq_f})]. Thus the excluded volume has a larger effect at larger
temperatures. On the other hand, also the Pauli-blocking gets weaker at larger
temperatures.

\begin{figure*}
\begin{center}
\includegraphics[width=0.75\columnwidth]{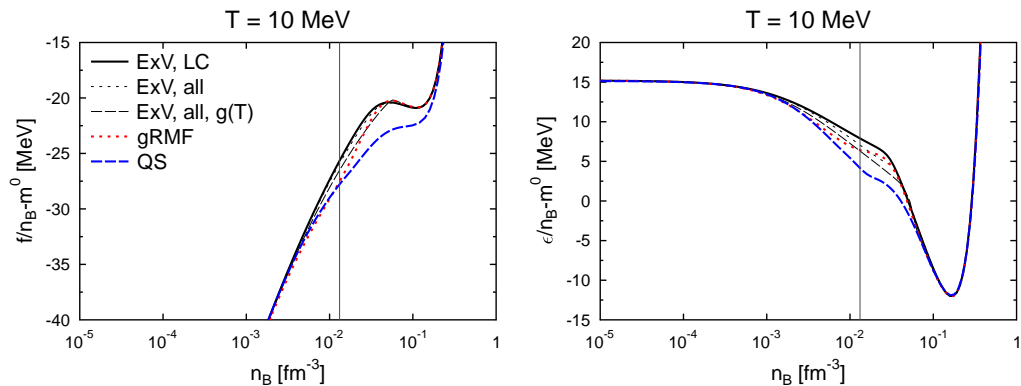}
\caption{\label{fig_thermo2}As in Fig.~\ref{fig_thermo1}, but now for $T=10$ MeV.}
\end{center}
\end{figure*}

The free energy and the internal energy of the models at $T=10$~MeV are compared
in  Fig.~\ref{fig_thermo2}. One sees that the ExV model with only light clusters
is now closer to the gRMF than to the QS model. However, compared to both of the
two models, the free and internal energies are increased in the ExV model at
almost all densities, even though the composition is similar. As noted before,
the contribution to the free energy which originates directly from the excluded
volume is proportional to $T$ and leads to an increase of the free energy.
However, the excluded volume does not give a direct contribution to the energy
density, which is also increased. We have a possible explanation for the observed
differences. Regarding the mean field, there are important conceptual differences
in the three models: In QS and gRMF all nucleons (bound in clusters and unbound)
contribute as sources for the meson fields. Furthermore, the light clusters are
treated as quasi-particles and acquire a mean-field self-energy. These effects
are absent in the ExV model, where the mean field is given only by the unbound
nucleons because the interacting nucleons are assumed to be outside of the light
clusters. There is no attraction acting on nuclei, but only the hard-core
repulsion. 

In the ExV calculation with all nuclei but without excited states (dotted thin
black lines in Figs.~\ref{fig_comp2} and \ref{fig_thermo2}), one finds that the
heavy nuclei for $T=10$~MeV are not as important as for $T=4$~MeV and that they
appear at larger densities. The maximum fractions of the light clusters are
reduced only slightly and the transition density to pure nucleon matter remains
similar. Here the tritons and helions behave similarly, and the effect of the
heavy nuclei observed in Fig.\ \ref{fig_comp1} is absent. At $T=10$ MeV the
inclusion of excited states (dashed thin black lines in Figs.~\ref{fig_comp2}
and \ref{fig_thermo2}) has a noticeable effect on the EOS and the composition.
The formation of heavy nuclei is statistically favored because of the large
number of accessible internal states. Their mass fraction exceeds 0.1 at $n_B
\sim 0.013$~fm$^{-3}$ and the maximum mass fraction is $\sim 0.55$ at $n_B \sim
0.052$~fm$^{-3}$. The internal partition function acts differently on the
abundances of the light clusters as can be seen by comparing the dotted with the
dashed thin black line in Fig.~\ref{fig_comp2}: the alpha-particle fraction is
increased most and the deuterons, on the other hand, show no visible change. The
deuteron is only very weakly bound with no bound excited state. Thus the
internal partition function remains small, in contrast to the much more strongly
bound alpha particle with a high cutoff for the maximum excitation energy. This
increase of the alpha-particle fraction is also present at very low densities.
These results might appear surprising, as the first excited state of the alpha
particle lies above 20~MeV. However, we also calculated the internal partition
function of the alpha particle using only the known excited states and found
that the expression of Eq.~(\ref{eq_gaz}) even underestimates the role of
excited states of the alpha particle for the temperatures investigated in this
article. In total there are 15 excited states between 20 and 30 MeV, with spins
of 0, 1, or 2 \cite{tilley92}, summing up to a notable contribution.

Similar as for $T=4$~MeV, the different models give different results for
densities larger than $ \sim 10^{-3}$ fm$^{-3}$. Even if we restrict the
comparison of the three models with only light nuclei at $T=10$~MeV to densities
to the left of the vertical lines, where heavy nuclei are not dominant, we still
observe some model dependency. Heavy nuclei appear at larger densities than at
$T=4$~MeV, so that the different predictions for light clusters of the three
models become relevant. For example, as depicted in Fig.\ \ref{fig_thermo2}, the
free energy per baryon differs up to a few MeV, and the internal energy per
baryon differs up to 5~MeV. The mass fractions of alphas and deuterons can be almost one
order of magnitude different, when reaching the vertical line.

\begin{figure*}
\begin{center}
\includegraphics[width=0.75\columnwidth]{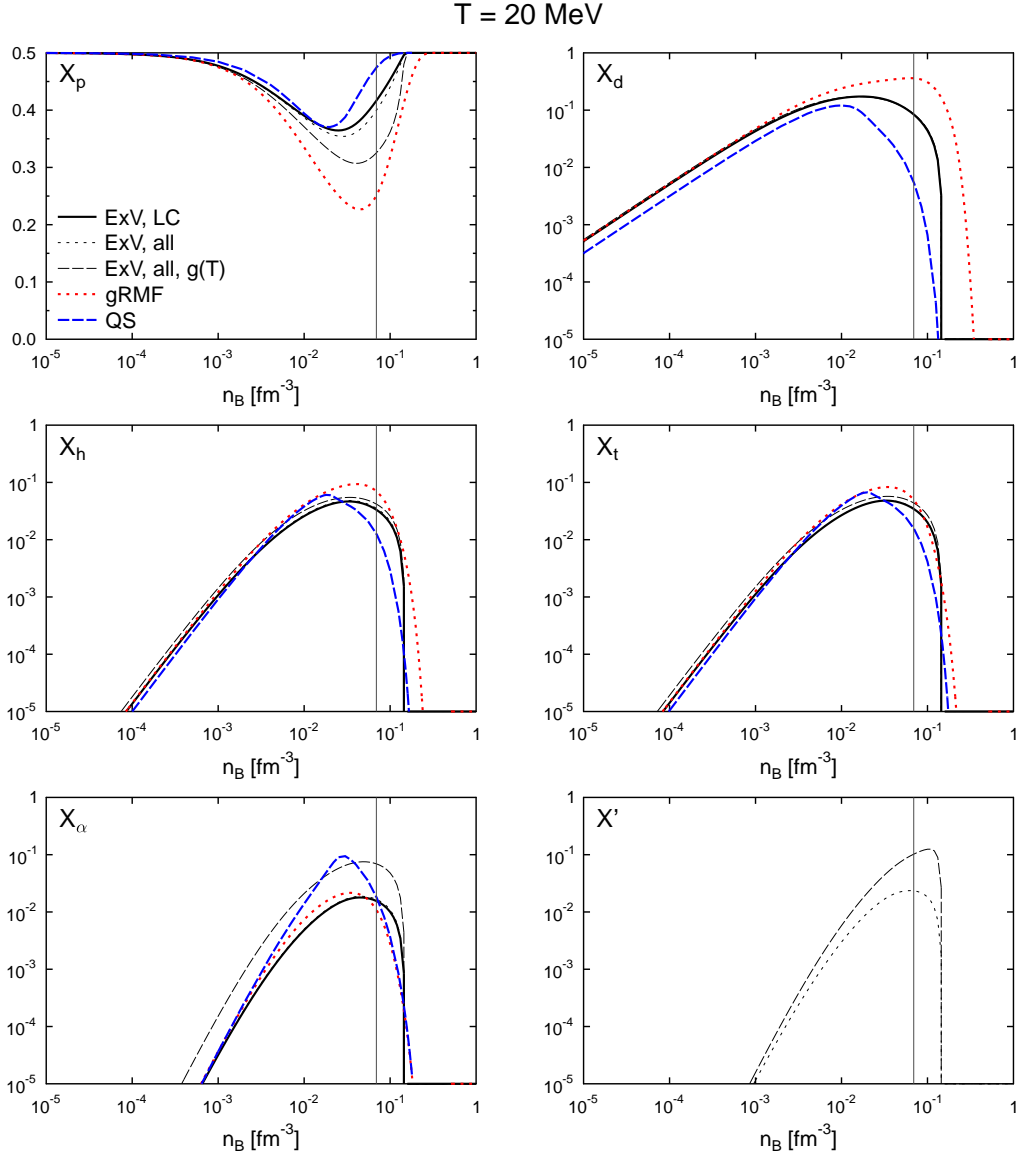}
\caption{\label{fig_comp3}As in Fig.~\ref{fig_comp1}, but now for $T=20$ MeV.}
 \end{center}
\end{figure*}

\begin{figure*}
\begin{center}
\includegraphics[width=0.75\columnwidth]{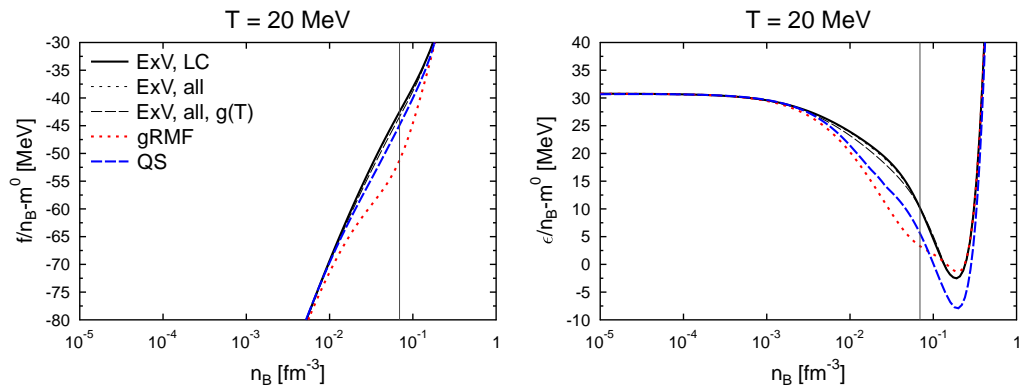}
\caption{\label{fig_thermo3}As in Fig.~\ref{fig_thermo1}, but now for $T=20$ MeV.}
 \end{center}
\end{figure*}

For $T=20$ MeV, which is shown in Fig.~\ref{fig_comp3}, the composition of the
ExV model evolves similarly with the density when compared with the results of  
the gRMF and QS models. The maximum mass fractions and the densities at which
the single clusters disappear are in a similar range. This supports the
conclusion about the effects of temperature which we have drawn before. The
reduced deuteron fraction in the QS model, in particular at low densities, is
now even more pronounced. As mentioned before, it is due to the more elaborated
treatment of the continuum states. The gRMF model, however, shows an increased
deuteron fraction as compared to the other approaches at higher densities before
the deuteron dissolves. Correspondingly, the proton fraction is reduced. There
are important differences in the EOS, which is shown in Fig.~\ref{fig_thermo3}, 
for densities where light clusters appear in large concentrations. All three
models exhibit rather different behaviors of the energies per nucleon. The ExV
model gives the largest energy and free energy, as for $T=10$~MeV. One also
finds that the QS model gives a significantly reduced internal energy compared
to the other two models close to saturation density. Fig.~\ref{fig_comp3} shows
that heavy nuclei play only a limited role at $T=20$ MeV, as there are only
small differences between the solid and the dotted black lines. If internal
excitations of the nuclei are considered in the partition function (dashed black
lines) the mass fraction of heavy nuclei increases above 0.1 at $n_B \sim
0.069$~fm$^{-3}$, but only in a narrow density region. Again, the alpha
particles profit the most from the inclusion of the excited states, as their
cutoff energy is very large. We remark that the direct contribution of the
internal partition function to the EOS (e.g., to the internal energy) is rather
small. This is not the case, if no cutoff in the integral for the excited states
was used: then arbitrary large excitation energies would give a contribution to
the energy density, which would therefore increase with increasing temperature
to unphysically large values.

Restricting the comparison of the light cluster models for $T= 20$~MeV to the
left of the vertical lines, there are pronounced differences for the three
models, due to the large densities involved. In the composition, the largest
differences are found between gRMF and QS. For the shown thermodynamic EOS
variables, gRMF and ExV give the most different results, deviating from each
other up to 10~MeV for the free energy and 5~MeV for the internal energy.

\section{Conclusions}
\label{sec_conc}

To include cluster formation in the nuclear matter equation of state, medium
effects have to be considered that suppress the abundances of light nuclei when
the baryon density approaches the saturation value. Basically, within a quantum
statistical approach, Pauli blocking as the consequence of anti-symmetrization
of fermionic wave functions is responsible for this suppression. Also well
known from atomic physics, the Pauli principle between fermions can be
considered as repulsion that is depicted by the excluded volume concept. The
excluded volume is an empirical parameter that, in simplest approximation, does
not depend on other parameter values such as density or temperature. This
concept is well proven to derive thermodynamic properties like the Van der Waals
EOS. We investigate the inclusion of medium effects to derive the nuclear matter
EOS with cluster formation by considering three different approaches: the
excluded volume (ExV) approach, the quantum statistical approach based on
many-particle theory, and the more challenging generalized relativistic
mean-field approach which starts from an effective Lagrangian, but superimposes
cluster formation as a first step in a semi-empirical way.

For the composition of dense matter with only light clusters, we conclude that
the excluded volume description can imitate the complicated quantum medium
effects relatively well at high temperatures. Similar mass fractions are reached
and the dissolution of light clusters happens at similar densities as in the
two quantum many-body approaches. In contrast, at low temperatures the ExV model
behaves very similar to an ideal gas and thus shows crucial deviations. The
better agreement at larger temperatures could partly be due to the excluded
volume term in the free-energy density, which is proportional to $T$ [see
Eq.~(\ref{eq_f})], and the reduced Pauli blocking. However, if one also
considers the results from the ExV model with heavy nuclei, one arrives at
partly different conclusions. If the comparison of the three light cluster
models is restricted to densities where the mass fraction of heavy nuclei is
below 0.1, the three models agree very well at low temperatures. The previously
mentioned crucial deviations of the ExV model at low temperatures are not
relevant, because they occur at densities where the composition is actually
dominated by heavy nuclei. For larger temperatures heavy nuclei appear at larger
densities, so that the slightly different model predictions for light clusters
are relevant for the full composition. Further uncertainty may come from the
inclusion of excited states which were seen to have a notable effect at large
temperatures.

For the EOS, i.e.\ the internal and free energy per baryon which were
investigated here, we found that in general it is not possible to correlate
thermodynamic variables like the energy density directly with the composition.
Even if two different models show a very similar density dependence of the
composition, the EOS can be notably different. For example, there seems to be a
systematic over-prediction of the internal and free energy with the ExV model
despite the good agreement for the mass fractions with the quantum many-body
models at large temperatures. We explained the increased energies by the missing
mean-field contribution of light nuclei in the ExV model. The free energy per
baryon and internal energy per baryon, respectively, differ for the three models
by up to 10 and 5~MeV, respectively, for the temperatures investigated here
and in the density region where light clusters and nucleons give the dominant
contribution.

Although the overall behavior of the mass fractions of light elements agrees
reasonably well for the three approaches investigated here, mass fractions of
individual light clusters can differ considerably. The accuracy to determine
mass fractions is related to the approximations treating the many-particle
problem, which become more intricate near the region where the bound states
disappear. Obviously, due to its phenomenological character, the excluded volume
concept can always only mimic the true quantum effects. A more sophisticated
quantum statistical approach has to reproduce not only the nuclear statistical
equilibrium in the low-density limit, but also the virial expansion. Therefore
the account of the contribution of continuum states, included in the QS
calculations, is essential, in particular at higher temperatures. Furthermore,
for the calculation of the mean fields instead of the approximation of
uncorrelated nucleons, correlations have to be included in a self-consistent
way (see \cite{roepke09}). Future progress in many-particle theory would improve
the accuracy to determine compositions or, more fundamentally, the spectral
function at higher densities.

For a more proper comparison with the ExV approach, the contribution of heavy
nuclei needs to be considered in the more sound but also more complicated
quantum many-body models. However, the results of the ExV approach for light
clusters are in satisfactory agreement with the other models. It is promising to
continue the investigations of the role of light clusters in core-collapse
supernovae and other dynamical astrophysical environments, mentioned in the
introduction, e.g., with one of the three models presented here. We conclude that
the excluded volume description could be used for such studies to identify the
principle effects of the light clusters, followed up with more elaborated models
like the quantum many-body models investigated here. 
  
\subsection*{Acknowledgments}
M.H.\ acknowledges support from the High Performance and High Productivity
Computing (HP2C) project. J.S.B.~is supported by the German Research Foundation
(DFG) within the framework of the excellence initiative through the Heidelberg
Graduate School of Fundamental Physics. This work has been  supported by the DFG
cluster of excellence ``Origin and Structure of the Universe'' and by CompStar a
research networking program of the European Science Foundation (ESF). M.H.\ is also
grateful for participating in the EuroGENESIS collaborative research program of
the ESF and the ENSAR/THEXO project.\\

\bibliographystyle{apsrev}
\bibliography{literature}

\begin{thebibliography}{29}
\expandafter\ifx\csname natexlab\endcsname\relax\def\natexlab#1{#1}\fi
\expandafter\ifx\csname bibnamefont\endcsname\relax
  \def\bibnamefont#1{#1}\fi
\expandafter\ifx\csname bibfnamefont\endcsname\relax
  \def\bibfnamefont#1{#1}\fi
\expandafter\ifx\csname citenamefont\endcsname\relax
  \def\citenamefont#1{#1}\fi
\expandafter\ifx\csname url\endcsname\relax
  \def\url#1{\texttt{#1}}\fi
\expandafter\ifx\csname urlprefix\endcsname\relax\def\urlprefix{URL }\fi
\providecommand{\bibinfo}[2]{#2}
\providecommand{\eprint}[2][]{\url{#2}}

\bibitem[{\citenamefont{{Natowitz} et~al.}(2010)\citenamefont{{Natowitz},
  {R{\"o}pke}, {Typel}, {Blaschke}, {Bonasera}, {Hagel}, {Kl{\"a}hn},
  {Kowalski}, {Qin}, {Shlomo} et~al.}}]{natowitz10}
\bibinfo{author}{\bibfnamefont{J.~B.} \bibnamefont{{Natowitz}}},
  \bibinfo{author}{\bibfnamefont{G.}~\bibnamefont{{R{\"o}pke}}},
  \bibinfo{author}{\bibfnamefont{S.}~\bibnamefont{{Typel}}},
  \bibinfo{author}{\bibfnamefont{D.}~\bibnamefont{{Blaschke}}},
  \bibinfo{author}{\bibfnamefont{A.}~\bibnamefont{{Bonasera}}},
  \bibinfo{author}{\bibfnamefont{K.}~\bibnamefont{{Hagel}}},
  \bibinfo{author}{\bibfnamefont{T.}~\bibnamefont{{Kl{\"a}hn}}},
  \bibinfo{author}{\bibfnamefont{S.}~\bibnamefont{{Kowalski}}},
  \bibinfo{author}{\bibfnamefont{L.}~\bibnamefont{{Qin}}},
  \bibinfo{author}{\bibfnamefont{S.}~\bibnamefont{{Shlomo}}},
  \bibnamefont{et~al.}, \bibinfo{journal}{Phys. Rev. Lett.}
  \textbf{\bibinfo{volume}{104}}, \bibinfo{pages}{202501}
  (\bibinfo{year}{2010}).

\bibitem[{\citenamefont{{Typel} et~al.}(2010)\citenamefont{{Typel},
  {R{\"o}pke}, {Kl{\"a}hn}, {Blaschke}, and {Wolter}}}]{typel09}
\bibinfo{author}{\bibfnamefont{S.}~\bibnamefont{{Typel}}},
  \bibinfo{author}{\bibfnamefont{G.}~\bibnamefont{{R{\"o}pke}}},
  \bibinfo{author}{\bibfnamefont{T.}~\bibnamefont{{Kl{\"a}hn}}},
  \bibinfo{author}{\bibfnamefont{D.}~\bibnamefont{{Blaschke}}},
  \bibnamefont{and} \bibinfo{author}{\bibfnamefont{H.~H.}
  \bibnamefont{{Wolter}}}, \bibinfo{journal}{Phys. Rev. C}
  \textbf{\bibinfo{volume}{81}}, \bibinfo{pages}{015803}
  (\bibinfo{year}{2010}).

\bibitem[{\citenamefont{{Qin} et~al.}(2011)\citenamefont{{Qin}, {Hagel},
  {Wada}, {Natowitz}, {Shlomo}, {Bonasera}, {Roepke}, {Typel}, {Chen}, {Huang}
  et~al.}}]{qin2011}
\bibinfo{author}{\bibfnamefont{L.}~\bibnamefont{{Qin}}},
  \bibinfo{author}{\bibfnamefont{K.}~\bibnamefont{{Hagel}}},
  \bibinfo{author}{\bibfnamefont{R.}~\bibnamefont{{Wada}}},
  \bibinfo{author}{\bibfnamefont{J.~B.} \bibnamefont{{Natowitz}}},
  \bibinfo{author}{\bibfnamefont{S.}~\bibnamefont{{Shlomo}}},
  \bibinfo{author}{\bibfnamefont{A.}~\bibnamefont{{Bonasera}}},
  \bibinfo{author}{\bibfnamefont{G.}~\bibnamefont{{Roepke}}},
  \bibinfo{author}{\bibfnamefont{S.}~\bibnamefont{{Typel}}},
  \bibinfo{author}{\bibfnamefont{Z.}~\bibnamefont{{Chen}}},
  \bibinfo{author}{\bibfnamefont{M.}~\bibnamefont{{Huang}}},
  \bibnamefont{et~al.}, \bibinfo{journal}{e-print arXiv:1110.3345}
  (\bibinfo{year}{2011}).

\bibitem[{\citenamefont{{R{\"u}ster} et~al.}(2006)\citenamefont{{R{\"u}ster},
  {Hempel}, and {Schaffner-Bielich}}}]{ruester06}
\bibinfo{author}{\bibfnamefont{S.~B.} \bibnamefont{{R{\"u}ster}}},
  \bibinfo{author}{\bibfnamefont{M.}~\bibnamefont{{Hempel}}}, \bibnamefont{and}
  \bibinfo{author}{\bibfnamefont{J.}~\bibnamefont{{Schaffner-Bielich}}},
  \bibinfo{journal}{Phys. Rev. C} \textbf{\bibinfo{volume}{73}},
  \bibinfo{pages}{035804} (\bibinfo{year}{2006}).

\bibitem[{\citenamefont{{Fischer} et~al.}(2011)\citenamefont{{Fischer},
  {Sagert}, {Pagliara}, {Hempel}, {Schaffner-Bielich}, {Rauscher},
  {Thielemann}, {K{\"a}ppeli}, {Mart{\'{\i}}nez-Pinedo}, and
  {Liebend{\"o}rfer}}}]{fischer11}
\bibinfo{author}{\bibfnamefont{T.}~\bibnamefont{{Fischer}}},
  \bibinfo{author}{\bibfnamefont{I.}~\bibnamefont{{Sagert}}},
  \bibinfo{author}{\bibfnamefont{G.}~\bibnamefont{{Pagliara}}},
  \bibinfo{author}{\bibfnamefont{M.}~\bibnamefont{{Hempel}}},
  \bibinfo{author}{\bibfnamefont{J.}~\bibnamefont{{Schaffner-Bielich}}},
  \bibinfo{author}{\bibfnamefont{T.}~\bibnamefont{{Rauscher}}},
  \bibinfo{author}{\bibfnamefont{F.-K.} \bibnamefont{{Thielemann}}},
  \bibinfo{author}{\bibfnamefont{R.}~\bibnamefont{{K{\"a}ppeli}}},
  \bibinfo{author}{\bibfnamefont{G.}~\bibnamefont{{Mart{\'{\i}}nez-Pinedo}}},
  \bibnamefont{and}
  \bibinfo{author}{\bibfnamefont{M.}~\bibnamefont{{Liebend{\"o}rfer}}},
  \bibinfo{journal}{Astrophys.\ J.\ Suppl.\ Series}
  \textbf{\bibinfo{volume}{194}}, \bibinfo{pages}{39} (\bibinfo{year}{2011}).

\bibitem[{\citenamefont{{Lattimer} and {Swesty}}(1991)}]{lattimer91}
\bibinfo{author}{\bibfnamefont{J.~M.} \bibnamefont{{Lattimer}}}
  \bibnamefont{and} \bibinfo{author}{\bibfnamefont{F.~D.} \bibnamefont{{Swesty}}}, 
\bibinfo{journal}{Nucl. Phys. A} \textbf{\bibinfo{volume}{535}},
  \bibinfo{pages}{331} (\bibinfo{year}{1991}).

\bibitem[{\citenamefont{{Shen} et~al.}(1998{\natexlab{a}})\citenamefont{{Shen},
  {Toki}, {Oyamatsu}, and {Sumiyoshi}}}]{shen98}
\bibinfo{author}{\bibfnamefont{H.}~\bibnamefont{{Shen}}},
  \bibinfo{author}{\bibfnamefont{H.}~\bibnamefont{{Toki}}},
  \bibinfo{author}{\bibfnamefont{K.}~\bibnamefont{{Oyamatsu}}},
  \bibnamefont{and}
  \bibinfo{author}{\bibfnamefont{K.}~\bibnamefont{{Sumiyoshi}}},
  \bibinfo{journal}{Nucl. Phys. A} \textbf{\bibinfo{volume}{637}},
  \bibinfo{pages}{435} (\bibinfo{year}{1998}{\natexlab{a}}).

\bibitem[{\citenamefont{{Shen} et~al.}(1998{\natexlab{b}})\citenamefont{{Shen},
  {Toki}, {Oyamatsu}, and {Sumiyoshi}}}]{shen98_2}
\bibinfo{author}{\bibfnamefont{H.}~\bibnamefont{{Shen}}},
  \bibinfo{author}{\bibfnamefont{H.}~\bibnamefont{{Toki}}},
  \bibinfo{author}{\bibfnamefont{K.}~\bibnamefont{{Oyamatsu}}},
  \bibnamefont{and}
  \bibinfo{author}{\bibfnamefont{K.}~\bibnamefont{{Sumiyoshi}}},
  \bibinfo{journal}{Prog. Theor. Phys.} \textbf{\bibinfo{volume}{100}},
  \bibinfo{pages}{1013} (\bibinfo{year}{1998}{\natexlab{b}}).

\bibitem[{\citenamefont{{Shen} et~al.}(2011{\natexlab{a}})\citenamefont{{Shen},
  {Horowitz}, and {Teige}}}]{shen2011a}
\bibinfo{author}{\bibfnamefont{G.}~\bibnamefont{{Shen}}},
  \bibinfo{author}{\bibfnamefont{C.~J.} \bibnamefont{{Horowitz}}},
  \bibnamefont{and} \bibinfo{author}{\bibfnamefont{S.}~\bibnamefont{{Teige}}},
  \bibinfo{journal}{\prc} \textbf{\bibinfo{volume}{83}},
  \bibinfo{pages}{035802} (\bibinfo{year}{2011}{\natexlab{a}}).

\bibitem[{\citenamefont{{Shen} et~al.}(2011{\natexlab{b}})\citenamefont{{Shen},
  {Horowitz}, and {O'Connor}}}]{shen2011b}
\bibinfo{author}{\bibfnamefont{G.}~\bibnamefont{{Shen}}},
  \bibinfo{author}{\bibfnamefont{C.~J.} \bibnamefont{{Horowitz}}},
  \bibnamefont{and}
  \bibinfo{author}{\bibfnamefont{E.}~\bibnamefont{{O'Connor}}},
  \bibinfo{journal}{\prc} \textbf{\bibinfo{volume}{83}},
  \bibinfo{pages}{065808} (\bibinfo{year}{2011}{\natexlab{b}}).

\bibitem[{\citenamefont{{Horowitz} and {Schwenk}}(2006)}]{horowitz06b}
\bibinfo{author}{\bibfnamefont{C.~J.} \bibnamefont{{Horowitz}}}
  \bibnamefont{and}
  \bibinfo{author}{\bibfnamefont{A.}~\bibnamefont{{Schwenk}}},
  \bibinfo{journal}{Nucl. Phys. A} \textbf{\bibinfo{volume}{776}},
  \bibinfo{pages}{55} (\bibinfo{year}{2006}).

\bibitem[{\citenamefont{{Hempel} and {Schaffner-Bielich}}(2010)}]{hempel10}
\bibinfo{author}{\bibfnamefont{M.}~\bibnamefont{{Hempel}}} \bibnamefont{and}
  \bibinfo{author}{\bibfnamefont{J.}~\bibnamefont{{Schaffner-Bielich}}},
  \bibinfo{journal}{Nucl. Phys. A} \textbf{\bibinfo{volume}{837}},
  \bibinfo{pages}{210} (\bibinfo{year}{2010}).

\bibitem[{\citenamefont{{Hempel} et~al.}(2011)\citenamefont{{Hempel},
  {Fischer}, {Schaffner-Bielich}, and {Liebend{\"o}rfer}}}]{hempel11}
\bibinfo{author}{\bibfnamefont{M.}~\bibnamefont{{Hempel}}},
  \bibinfo{author}{\bibfnamefont{T.}~\bibnamefont{{Fischer}}},
  \bibinfo{author}{\bibfnamefont{J.}~\bibnamefont{{Schaffner-Bielich}}},
  \bibnamefont{and}
  \bibinfo{author}{\bibfnamefont{M.}~\bibnamefont{{Liebend{\"o}rfer}}},
  \bibinfo{journal}{e-print arXiv:1108.0848}  (\bibinfo{year}{2011}).

\bibitem[{\citenamefont{{Sumiyoshi} and {R{\"o}pke}}(2008)}]{sumiyoshi08}
\bibinfo{author}{\bibfnamefont{K.}~\bibnamefont{{Sumiyoshi}}} \bibnamefont{and}
  \bibinfo{author}{\bibfnamefont{G.}~\bibnamefont{{R{\"o}pke}}},
  \bibinfo{journal}{Phys. Rev. C} \textbf{\bibinfo{volume}{77}},
  \bibinfo{pages}{055804} (\bibinfo{year}{2008}).

\bibitem[{\citenamefont{{R{\"o}pke}}(2009)}]{roepke09}
\bibinfo{author}{\bibfnamefont{G.}~\bibnamefont{{R{\"o}pke}}},
  \bibinfo{journal}{Phys. Rev. C} \textbf{\bibinfo{volume}{79}},
  \bibinfo{pages}{014002} (\bibinfo{year}{2009}).

\bibitem[{\citenamefont{{O'Connor} et~al.}(2007)\citenamefont{{O'Connor},
  {Gazit}, {Horowitz}, {Schwenk}, and {Barnea}}}]{oconnor07}
\bibinfo{author}{\bibfnamefont{E.}~\bibnamefont{{O'Connor}}},
  \bibinfo{author}{\bibfnamefont{D.}~\bibnamefont{{Gazit}}},
  \bibinfo{author}{\bibfnamefont{C.~J.} \bibnamefont{{Horowitz}}},
  \bibinfo{author}{\bibfnamefont{A.}~\bibnamefont{{Schwenk}}},
  \bibnamefont{and} \bibinfo{author}{\bibfnamefont{N.}~\bibnamefont{{Barnea}}},
  \bibinfo{journal}{Phys. Rev. C} \textbf{\bibinfo{volume}{75}},
  \bibinfo{pages}{055803} (\bibinfo{year}{2007}).

\bibitem[{\citenamefont{{Arcones} et~al.}(2008)\citenamefont{{Arcones},
  {Mart{\'{\i}}nez-Pinedo}, {O'Connor}, {Schwenk}, {Janka}, {Horowitz}, and
  {Langanke}}}]{arcones08}
\bibinfo{author}{\bibfnamefont{A.}~\bibnamefont{{Arcones}}},
  \bibinfo{author}{\bibfnamefont{G.}~\bibnamefont{{Mart{\'{\i}}nez-Pinedo}}},
  \bibinfo{author}{\bibfnamefont{E.}~\bibnamefont{{O'Connor}}},
  \bibinfo{author}{\bibfnamefont{A.}~\bibnamefont{{Schwenk}}},
  \bibinfo{author}{\bibfnamefont{H.-T.} \bibnamefont{{Janka}}},
  \bibinfo{author}{\bibfnamefont{C.~J.} \bibnamefont{{Horowitz}}},
  \bibnamefont{and}
  \bibinfo{author}{\bibfnamefont{K.}~\bibnamefont{{Langanke}}},
  \bibinfo{journal}{Phys. Rev. C} \textbf{\bibinfo{volume}{78}},
  \bibinfo{pages}{015806} (\bibinfo{year}{2008}).

\bibitem[{\citenamefont{{Heckel} et~al.}(2009)\citenamefont{{Heckel},
  {Schneider}, and {Sedrakian}}}]{heckel09}
\bibinfo{author}{\bibfnamefont{S.}~\bibnamefont{{Heckel}}},
  \bibinfo{author}{\bibfnamefont{P.~P.} \bibnamefont{{Schneider}}},
  \bibnamefont{and}
  \bibinfo{author}{\bibfnamefont{A.}~\bibnamefont{{Sedrakian}}},
  \bibinfo{journal}{Phys. Rev. C} \textbf{\bibinfo{volume}{80}},
  \bibinfo{pages}{015805} (\bibinfo{year}{2009}).

\bibitem[{\citenamefont{{R{\"o}pke}}(2011)}]{roepke11a}
\bibinfo{author}{\bibfnamefont{G.}~\bibnamefont{{R{\"o}pke}}},
  \bibinfo{journal}{Nucl. Phys. A} \textbf{\bibinfo{volume}{867}},
  \bibinfo{pages}{66} (\bibinfo{year}{2011}).

\bibitem[{\citenamefont{{Typel}}(2005)}]{typel05}
\bibinfo{author}{\bibfnamefont{S.}~\bibnamefont{{Typel}}},
  \bibinfo{journal}{Phys. Rev. C} \textbf{\bibinfo{volume}{71}},
  \bibinfo{pages}{064301} (\bibinfo{year}{2005}).

\bibitem[{\citenamefont{{Audi} et~al.}(2003)\citenamefont{{Audi}, {Wapstra},
  and {Thibault}}}]{AudiWapstra}
\bibinfo{author}{\bibfnamefont{G.}~\bibnamefont{{Audi}}},
  \bibinfo{author}{\bibfnamefont{A.~H.} \bibnamefont{{Wapstra}}},
  \bibnamefont{and}
  \bibinfo{author}{\bibfnamefont{C.}~\bibnamefont{{Thibault}}},
  \bibinfo{journal}{Nucl. Phys. A} \textbf{\bibinfo{volume}{729}},
  \bibinfo{pages}{337} (\bibinfo{year}{2003}).

\bibitem[{\citenamefont{{M{\"o}ller} et~al.}(1997)\citenamefont{{M{\"o}ller},
  {Nix}, and {Kratz}}}]{Moller97}
\bibinfo{author}{\bibfnamefont{P.}~\bibnamefont{{M{\"o}ller}}},
  \bibinfo{author}{\bibfnamefont{J.~R.} \bibnamefont{{Nix}}}, \bibnamefont{and}
  \bibinfo{author}{\bibfnamefont{K.-L.} \bibnamefont{{Kratz}}},
  \bibinfo{journal}{Atom. Data Nucl. Data Tables}
  \textbf{\bibinfo{volume}{66}}, \bibinfo{pages}{131} (\bibinfo{year}{1997}).

\bibitem[{\citenamefont{{Nadyozhin} and {Yudin}}(2005)}]{nadyozhin05}
\bibinfo{author}{\bibfnamefont{D.~K.} \bibnamefont{{Nadyozhin}}}
  \bibnamefont{and} \bibinfo{author}{\bibfnamefont{A.~V.}
  \bibnamefont{{Yudin}}}, \bibinfo{journal}{Astron. L.}
  \textbf{\bibinfo{volume}{31}}, \bibinfo{pages}{271} (\bibinfo{year}{2005}).

\bibitem[{\citenamefont{{Potekhin} et~al.}(2009)\citenamefont{{Potekhin},
  {Chabrier}, {Chugunov}, {DeWitt}, and {Rogers}}}]{potekhin09}
\bibinfo{author}{\bibfnamefont{A.~Y.} \bibnamefont{{Potekhin}}},
  \bibinfo{author}{\bibfnamefont{G.}~\bibnamefont{{Chabrier}}},
  \bibinfo{author}{\bibfnamefont{A.~I.} \bibnamefont{{Chugunov}}},
  \bibinfo{author}{\bibfnamefont{H.~E.} \bibnamefont{{DeWitt}}},
  \bibnamefont{and} \bibinfo{author}{\bibfnamefont{F.~J.}
  \bibnamefont{{Rogers}}}, \bibinfo{journal}{Phys. Rev. E}
  \textbf{\bibinfo{volume}{80}}, \bibinfo{pages}{047401}
  (\bibinfo{year}{2009}).

\bibitem[{\citenamefont{{Potekhin} and {Chabrier}}(2010)}]{potekhin10}
\bibinfo{author}{\bibfnamefont{A.~Y.} \bibnamefont{{Potekhin}}}
  \bibnamefont{and}
  \bibinfo{author}{\bibfnamefont{G.}~\bibnamefont{{Chabrier}}},
  \bibinfo{journal}{Contrib. Plasma Phys.} \textbf{\bibinfo{volume}{50}},
  \bibinfo{pages}{82} (\bibinfo{year}{2010}).

\bibitem[{\citenamefont{{F{\'a}i} and {Randrup}}(1982)}]{fai82}
\bibinfo{author}{\bibfnamefont{G.}~\bibnamefont{{F{\'a}i}}} \bibnamefont{and}
  \bibinfo{author}{\bibfnamefont{J.}~\bibnamefont{{Randrup}}},
  \bibinfo{journal}{Nucl. Phys. A} \textbf{\bibinfo{volume}{381}},
  \bibinfo{pages}{557} (\bibinfo{year}{1982}).

\bibitem[{\citenamefont{{R\"opke}}(1984)}]{roepke84}
\bibinfo{author}{\bibfnamefont{G.}~\bibnamefont{{R\"opke}}},
  \bibinfo{journal}{Wiss.\ Z.\ Univ.\ Rostock} \textbf{\bibinfo{volume}{33}}, \bibinfo{pages}{33}
  (\bibinfo{year}{1984}).

\bibitem[{\citenamefont{{Yudin}}(2011)}]{yudin11}
\bibinfo{author}{\bibfnamefont{A.~V.} \bibnamefont{{Yudin}}},
  \bibinfo{journal}{Astronomy Letters} \textbf{\bibinfo{volume}{37}},
  \bibinfo{pages}{576} (\bibinfo{year}{2011}).

\bibitem[{\citenamefont{{Tilley} et~al.}(1992)\citenamefont{{Tilley}, {Weller},
  and {Hale}}}]{tilley92}
\bibinfo{author}{\bibfnamefont{D.~R.} \bibnamefont{{Tilley}}},
  \bibinfo{author}{\bibfnamefont{H.~R.} \bibnamefont{{Weller}}},
  \bibnamefont{and} \bibinfo{author}{\bibfnamefont{G.~M.}
  \bibnamefont{{Hale}}}, \bibinfo{journal}{Nucl. Phys. A}
  \textbf{\bibinfo{volume}{541}}, \bibinfo{pages}{1} (\bibinfo{year}{1992}).

\end{thebibliography}

\end{document}